\documentclass[12pt,english]{article}
\usepackage{mathptmx}
\usepackage[T1]{fontenc}
\usepackage[latin9]{inputenc}
\usepackage[letterpaper]{geometry}
\usepackage{color}
\usepackage{float}
\usepackage{amsmath}
\usepackage{amsthm}
\usepackage{graphicx}
\usepackage[authoryear]{natbib}

\makeatletter
\newcommand{\lyxaddress}[1]{
\par {\raggedright #1
\noindent\par}
}

\date{}
\usepackage{xfrac}

\makeatother

\usepackage{babel}
\begin{document}

\title{Wavemaker theories for acoustic-gravity waves over a finite depth}

\author{Miao Tian$^{1,2}$, Usama Kadri$^{3,4}$}
\maketitle
\lyxaddress{1. Department of Physical Oceanography, Woods Hole Oceanographic
Institute, Woods Hole, MA 02543, USA. }
\lyxaddress{2. The Hatter Department of Marine Technologies, University of Haifa, 
Haifa 3498838.}
\lyxaddress{{3. School of Mathematics, Cardiff University, Cardiff, CF24 4AG, UK. }}
\lyxaddress{4. Department of Mathematics, Massachusetts Institute of Technology,
Cambridge, MA 02139, USA.}

\begin{abstract}
Acoustic-gravity waves (hereafter AGWs) in ocean have received much
interest recently, mainly with respect to early detection of tsunamis
as they travel at near the speed of sound in water which makes them
ideal candidates for early detection of tsunamis. While the generation
mechanisms of AGWs have been studied from the perspective of vertical
oscillations of seafloor and triad wave-wave interaction, in the current
study we are interested in their generation by wave-structure interaction
with possible implication to the energy sector. Here, we develop two
wavemaker theories to analyze different wave modes generated by impermeable
(the classic Havelock's theory) and porous (porous wavemaker theory)
plates in weakly compressible fluids. Slight modification has been
made to the porous theory so that, unlike the previous theory, the
new solution depends on the geometry of the plate. The expressions
for three different types of plates (piston, flap, delta-function)
are introduced. Analytical solutions are also derived for the potential
amplitude of the gravity, acoustic-gravity, evanescent waves, as well
as the surface elevation, velocity distribution, and pressure for
AGWs. Both theories reduce to previous results for incompressible
flow when the compressibility is neglected. We also show numerical
examples for AGWs generated in a wave flume as well as in deep ocean.
Our current study sets the theoretical background towards remote sensing
by AGWs, for optimized deep ocean wave-power harnessing, among others.
\end{abstract}

\section{Introduction}
Wavemaker theory has received increasing attention not only because
its feasibility on generating waves in laboratory experiments, but
also due to its application in design of wave-energy harnessing devices
\citep{Mei_PTRS_2012}. The classic problem of surface waves generated
by a wavemaker in infinitely deep ocean was investigated {by
Havelock, as in Ref. \citet{Havelock_1929},} and later extended to
the case of finite water depth \citep[e.g.,][]{Ursel_JFM1960}. The
wavemaker was treated as a vertical impermeable plate which oscillates
horizontally and periodically with a small displacement, and the fluid
was assumed incompressible. In all these formulations the wave motion
was governed by linear wave theory. Extensions to a directional wavemaker
problem with slowly-varying depth can be found in \citet{Dalrymple_1989}. 

The impermeability of the plate is unrealistic for a plate in a wave
flume, not to mention a landslide in deep ocean. \citet{Madsen_1970}
examined the influence of leakage around the wavemaker on the wave
amplitude and concluded that the porous effect can largely reduce
the wave amplitude. Therefore it would be more appropriate to take
porosity effects into account for many applications.

Recently, acoustic-gravity waves (hereafter AGWs) in a compressible
ocean have received much interest, because AGWs travel significantly
faster than the tsunami, and become ideal candidates for early detection
of tsunami by the use of bottom-pressure records \citep{Stiassnie_2010,HendinStiassnie_POF2013}.
AGWs {can} interact with continental shelves \citep{Kadri_Stiassnie_2012}, {ice-sheets \citep{Kadri_ice_2016}},
and \text{blue}{might} be responsible for deep-ocean \text{blue}{water transportation and} circulation \citep{Kadri_2014}.
In contrast to the decaying vertical structure of gravity-wave modes,
{the wave amplitudes of AGWs exhibit sinusoidal variation
in the vertical direction.} {Therefore wave-energy
harnessing devices that are placed in deep water (where the decaying
gravity wave modes vanish) can potentially make use of AGWs, because
the induced measurable pressure signature may reach a maximum at the seabed. } {While harnessing energy of AGWs might become possible in the future, e.g. based on a triad interaction mechanism similar to that presented by \cite{Kadri_EJM2016} or \cite{Kadri_Akylas_JFM2016}, a more immediate application is the detection of sea-state in wave harnessing farms. Here, we show that AGWs radiate by the harnessing devices, or namely wavemakers, carrying information on their source at the speed of sound in water. To this end, w}hile the generation
mechanisms of AGWs have been studied from the perspective of vertical
oscillations of seafloor \citep{Yamamoto_1982,Stiassnie_2010, Oliveira_2016} and
triad wave-wave interaction \citep{Kadri_JFM2013,Kadri_EJM2015,Kadri_EJM2016,Kadri_Akylas_JFM2016},
here we are {particularly} interested in their generation by horizontally-moving wavemakers.

We develop Havelock's and porous wavemaker theories
for weakly compressible fluids. The paper is organized as follows:
in section \ref{sec:Governing-equations} the problem is formulated
with the governing equations, while the wavemakers is treated as along-channel
boundary conditions. The general solution to the governing equations
is presented in Section \ref{sec:General-Solution}, followed by the
Havelock's and porous-wavemaker solutions in Section \ref{sec:Wavemaker-problem}.
Section \ref{sec:Examples} presents examples for three types of wavemaker
placed in a wave flume as well as in deep ocean. The work is summarized
in Section \ref{sec:Conclusion}.

\section{Governing equations \label{sec:Governing-equations}}
We take $x$ and $z$ the horizontal and cal coordinate respectively,
and consider a wavemaker with its plate initially located at $x=0$.
The wavemaker oscillates horizontally along the $x$-axis with a displacement
$s_{0}$ {given by}
\begin{equation}
s_{0}\left(x,z,t\right)=d\left(z\right)\exp\left(-\mbox{i}\omega t\right), \quad d\ll h \label{eq:piston motion}
\end{equation}
where $\omega$ is the radian frequency, $d\left(z\right)$ is the
maximum amplitude of oscillation, assumed to be much smaller than
the undisturbed fluid depth $h$, and $t$ is the time. The horizontal
velocity and acceleration of the wavemaker are
\begin{equation}
u_{0}=-\mbox{i}\omega d\exp\left(-\mbox{i}\omega t\right),\quad a_{0}=-\omega^{2}d\exp\left(-\mbox{i}\omega t\right).\label{eq:piston-velocity-acceleration}
\end{equation}

The equation that governs the irrotational motions of acoustic-gravity
waves throughout the entire water column is 
\begin{equation}
\Phi_{tt}=c^{2}\left(\Phi_{xx}+\Phi_{zz}\right),\quad-h<z<\eta,\label{eq:GoverningEq}
\end{equation}
where $\Phi$ is the velocity potential, and $c$ is the speed of
sound in water. The linearized kinematic and dynamic conditions at
the free surface are
\begin{alignat}{1}
\Phi_{z} & =\eta_{t},\quad z=\eta,\label{eq:kinematic}\\
\Phi_{t}+g\eta & =0,\quad z=\eta,\label{eq:dynamical}
\end{alignat}
where $\eta$ is the free surface elevation. Expanding \eqref{eq:kinematic}
and \eqref{eq:dynamical} at $z=0$ and eliminating $\eta$ yield
the approximated surface boundary condition {(e.g.,
\citealt{Lamb-book})}
\begin{equation}
\Phi_{tt}+g\Phi_{z}=0,\quad z=0.\label{eq:FreeSurface-BC}
\end{equation}

Finally the kinematic bottom boundary condition for a flat bottom
is given by
\begin{equation}
\Phi_{z}=0,\quad z=-h,\label{eq:Bottom-BC}
\end{equation}
which indicates the vertical velocity of the fluid must be zero at
the bottom.

Equations \eqref{eq:GoverningEq}, \eqref{eq:FreeSurface-BC}, and
\eqref{eq:Bottom-BC} formulate the linear problem of water wave propagation
over a finite depth in a weakly compressible fluid. Appropriate along-channel
boundary conditions depending on the types of wavemaker can be included
{to define the problem completely.} 

For the classic Havelock's wave-maker theory \citep[e.g.,][]{Havelock_1929,Ursel_JFM1960,Stuhlmeier_2015},
the boundary condition is
\begin{equation}
\Phi_{x}=u_{0},\;x=0\label{eq:Wavemaker-BC-Havelock}
\end{equation}
where $u_{0}$ is the horizontal velocity of the stroke motion.

For a porous-wavemaker problem, the boundary condition at the wavemaker
is given by {Chwang \citet{Chwang_JFM1983}}. The
hydrodynamic pressure $p\left(x,z,t\right)$ is associated with the
velocity potential $\Phi$ via the linearized Bernoulli equation as
\begin{equation}
p=-\rho\Phi_{t}\label{eq:P-Phi}
\end{equation}
{in which} $\rho$ is the water density. 

The pressure on the positive and negative sides of the wavemaker are
related as
\begin{equation}
p\left(0,z,t\right)=p^{+}\left(z,t\right)=-p^{-}\left(z,t\right).\label{eq:P+_P-}
\end{equation}

The normal velocity towards the porous plate is equal to the velocity
of the stroke motion $u_{0}$, which is linearly proportional to the
pressure difference between the two sides of the wavemaker \citep{Taylor_1956},
{so that}
\begin{equation}
u_{0}\left(z,t\right)=\frac{2b}{\mu}p\left(0,z,t\right)\label{eq:W-P}
\end{equation}
{Here, }$\mu$ is the dynamic viscosity, and $b$
is the coefficient which represents the width of the plate and has
the dimension of a length.

\section{General Solution \label{sec:General-Solution}}
In accordance with the periodic motion of the wavemaker, $\Phi$, $\eta$,
and $p$ are assumed to be periodic functions in $t$ with a time
factor $\exp\left(-\mbox{i}\omega t\right)$, i.e.
\begin{equation}
\Phi=\phi\left(x,z\right)\exp\left(-\mbox{i}\omega t\right),\quad \eta=a\left(x\right)\exp\left(-\mbox{i}\omega t\right),\quad p=p\left(x,z\right)\exp\left(-\mbox{i}\omega t\right).\label{eq:exp-t}
\end{equation}

Using \eqref{eq:exp-t}, equation \eqref{eq:GoverningEq} reduces
to 
\begin{equation}
\phi_{xx}+\phi_{zz}+k_{c}^{2}\phi=0,\quad k_{c}=\omega/c,\label{eq:Helmholtz}
\end{equation}
where $k_{c}$ is a compressibility coefficient. 
{Similarly, substiting equation \eqref{eq:exp-t}
into equations \eqref{eq:FreeSurface-BC} and \eqref{eq:Bottom-BC}
yields the boundary conditions in terms of $\phi$,
\begin{equation}
-\omega^{2}\phi+g\phi_{z}=0,\quad z=0;\label{eq:FreeSurface-BC-Fourier}
\end{equation}
}
{
\begin{equation}
\phi_{z}=0,\quad z=-h.\label{eq:Bottom-BC-Fourier}
\end{equation}
Following similar steps as in \citet{Stiassnie_2010} and
\citet{Yamamoto_1982} the solution of equations \eqref{eq:Helmholtz}-\eqref{eq:Bottom-BC-Fourier}
is obtained, }
{
\begin{alignat}{1}
\phi & =A_{0}\exp\left(\mbox{i}k_{0}x\right)\cosh\left(\lambda_{0}\left(z+h\right)\right)\nonumber \\
 & +\sum_{n=1}^{N}A_{n}\exp\left(\mbox{i}k_{n}x\right)\cos\left(\lambda_{n}\left(z+h\right)\right)\nonumber \\
 & +\sum_{n=N+1}^{\infty}B_{n}\exp\left(-\kappa_{n}x\right)\cos\left(\lambda_{n}\left(z+h\right)\right),\label{eq:final solution-finite}
\end{alignat}
}
{Here, $\lambda_{0}$ and $\lambda_{n}$ (real and
positive) are the solutions of}
{
\begin{gather}
\omega^{2}=g\lambda_{0}\tanh\left(\lambda_{0}h\right);\label{eq:dispersion-gravity}\\
\omega^{2}=-g\lambda_{n}\tan\left(\lambda_{n}h\right),\quad n=1,2,3,\ldots,\label{eq:dispersion-AGW-Evanescent}
\end{gather}
}
{where $\lambda_{n}$ is the $n$-th eigenvalue and
$n$ is the mode number. With specified $\omega$ and $h$, equation
\eqref{eq:dispersion-gravity} has one real solution for $\lambda_{0}$;
while equation \eqref{eq:dispersion-AGW-Evanescent} involves involves
infinitely-many different $\lambda_{n}$.}

{Here, $k_{0}$, $k_{n}$, $\kappa_{n}$ (real and
positive) are given by
\begin{equation}
k_{0}=\sqrt{k_{c}^{2}+\lambda_{0}^{2}},\label{eq:k0}
\end{equation}
\begin{equation}
k_{n}=\sqrt{k_{c}^{2}-\lambda_{n}^{2}},\quad n=1,2,...,N;\quad k_{c}>\lambda_{N},\label{eq:kn}
\end{equation}
\begin{equation}
\kappa_{n}=\sqrt{\lambda_{n}^{2}-k_{c}^{2}},\quad n=N+1,...;\quad k_{c}<\lambda_{N+1}\label{eq:kappa-n}
\end{equation}
}
{where $N$ in represents the nearest integer smaller
than $[\frac{\omega h}{\pi c}+\frac{1}{2}]$, as in Ref. \citep{Kadri_Stiassnie_2012}.
The three terms on the right-hand-side of equation \eqref{eq:final solution-finite}
represent the gravity, acoustic-gravity, and evanescent modes, respectively.}

\section{Wavemaker problem\label{sec:Wavemaker-problem}}
\subsection{Solution for Havelock's wavemaker}
{Since $\cosh\left(\lambda_{0}\left(z+h\right)\right)$
and $\cos\left(\lambda_{n}\left(z+h\right)\right)$in equation \eqref{eq:final solution-finite}
are the eigenfunctions of the boundary value problem in $z$, they
are orthogonal over the interval from $z=0$ to $z=-h$ based on the
Sturm-Liouville theory. Therefore we substitute equations \eqref{eq:exp-t},
and \eqref{eq:final solution-finite} into equation \eqref{eq:Wavemaker-BC-Havelock},
multiply by $\cosh\left(\lambda_{0}\left(z+h\right)\right)$ and $\cos\left(\lambda_{n}\left(z+h\right)\right)$
and integrate over the water column from $z=-h$ to $z=0$ so that $A_{0}$,
$A_{n}$, $B_{n}$ can be calculated as }
\begin{alignat}{1}
A_{0} & =\frac{-2\omega}{h\sqrt{k_{c}^{2}+\lambda_{0}^{2}}\left(1+CQ_{0}^{2}\right)}\int\limits_{-h}^{0}d\cosh\left(\lambda_{0}\left(z+h\right)\right)\mbox{d}z,\label{eq:A0-Havelock}\\
A_{n} & =\frac{-2\omega}{h\sqrt{k_{c}^{2}-\lambda_{n}^{2}}\left(1-CQ_{n}^{2}\right)}\int\limits_{-h}^{0}d\cos\left(\lambda_{n}\left(z+h\right)\right)\mbox{d}z,\label{eq:A-Havelock}\\
B_{n} & =\frac{2\mbox{i}\omega}{h\sqrt{\lambda_{n}^{2}-k_{c}^{2}}\left(1-CQ_{n}^{2}\right)}\int\limits_{-h}^{0}d\cos\left(\lambda_{n}\left(z+h\right)\right)\mbox{d}z,\label{eq:B-Havelock}
\end{alignat}
where
\begin{alignat}{1}
Q_{0} & =\sinh\lambda_{0}h,\quad Q_{n}=\sin\lambda_{n}h,\quad C=\frac{g}{\omega^{2}h}.\label{eq:Q0-Qn-C}
\end{alignat}

In the incompressible case ($c\rightarrow\infty$), equations \eqref{eq:A0-Havelock}
and \eqref{eq:B-Havelock} are essentially the same as the solutions
for gravity and evanescence modes in \citet{Dean_Dalrymple_1991}
(equations (6.21) and (6.22)). The extra term $A_{n}$ comes from
the newly-generated AGW mode due to the compressibility of the fluid. 

\subsection{Solution for porous wavemaker}
{Following similar steps using \eqref{eq:W-P} together
with \eqref{eq:P-Phi}, \eqref{eq:exp-t}, and \eqref{eq:final solution-finite}
and the orthogonality of $\cosh\left(\lambda_{0}\left(z+h\right)\right)$
and $\cos\left(\lambda_{n}\left(z+h\right)\right)$, we can derive
expressions of $A_{0}$, $A_{n}$, $B_{n}$ for a porous wavemaker.
\citet{Chwang_JFM1983} described a similar problem for incompressible
flow and derived the solutions. Chwang's solution, however, indicates
that the produced waves have the same amplitudes regardless of the
geometry of the plate. In order to consider different plate types,
we modified Chwang's method and derive an alternative solution in
a similar form as the Havelock's \citep{Dean_Dalrymple_1991}}
\begin{alignat}{1}
A_{0} & =G_{0}\frac{-2\omega}{h\sqrt{k_{c}^{2}+\lambda_{0}^{2}}\left(1+CQ_{0}^{2}\right)}\int\limits_{-h}^{0}d\cosh\left(\lambda_{0}\left(z+h\right)\right)\mbox{d}z,\label{eq:A0-Porous}\\
A_{n} & =G_{n}\frac{-2\omega}{h\sqrt{k_{c}^{2}-\lambda_{n}^{2}}\left(1-CQ_{n}^{2}\right)}\int\limits_{-h}^{0}d\cos\left(\lambda_{n}\left(z+h\right)\right)\mbox{d}z,\label{eq:A-Porous}\\
B_{n} & =H_{n}\frac{2\mbox{i}\omega}{h\sqrt{\lambda_{n}^{2}-k_{c}^{2}}\left(1-CQ_{n}^{2}\right)}\int\limits_{-h}^{0}d\cos\left(\lambda_{n}\left(z+h\right)\right)\mbox{d}z,\label{eq:B-Porous}
\end{alignat}
where the porous factors are
\begin{gather}
\begin{split}
G_{0}=\frac{\mu\sqrt{k_{c}^{2}+\lambda_{0}^{2}}}{2\omega\rho b},\quad G_{n}=\frac{\mu\sqrt{k_{c}^{2}-\lambda_{n}^{2}}}{2\omega\rho b},\quad n=1,2,\ldots N,\\H_{n}=\frac{-\mbox{i}\mu\sqrt{\lambda_{n}^{2}-k_{c}^{2}}}{2\omega\rho b}\quad n=N+1,N+2,\ldots.\label{eq:porous factor}
\end{split}
\end{gather}

We focus on the porous factor $G_{n}$ that is associated with AGW
modes. As pointed out in \citet{Chwang_JFM1983}, the reciprocal of
$G_{n}$ in equation \eqref{eq:porous factor} can be understood as
a Reynolds number for the flow passing through the porous wavemaker,
while $G_{n}$ also measures the {porosity}. For example,
$G_{n}=0$ (or equivalently, $\mu=0$) means the wavemaker is completely
permeable. Obviously, the expressions reduce to the Havelock's solution
when the porous factors, $G_{0}$, $G_{n}$, $H_{n}$, are unity.
Moreover, as $\lambda_{n}$ increases with mode number $n$, the porous
factor $G_{n}$ decreases for higher AGW modes, meaning that the porous
media dissipates more energy from shorter waves (lower modes). Specifying
values for the porous factor of the gravity mode $G_{0}$ ($0.1$,
$0.2$, $0.5,$ etc.), \citet{Chwang_JFM1983} showed surface elevation
of gravity waves produced by the wavemaker. In this study, the porous
factor for the first AGW mode $G_{1}$ will be set to $0.5$ for illustration
purposes, while $G_{n}$ $\left(n=2,3,\ldots\right)$ can be determined
based on that. 

The amplitudes of the surface elevation, horizontal velocity, and
pressure at AGW mode $n$ can also be given {employing} equations \eqref{eq:dynamical},
\eqref{eq:P-Phi}, \eqref{eq:exp-t}, \eqref{eq:final solution-finite}
\begin{alignat}{1}
a_{n} & =\mbox{i}\frac{\omega}{g}A_{n}\exp\left(\mbox{i}\sqrt{k_{c}^{2}-\lambda_{n}^{2}}x\right)\cos\left(\lambda_{n}h\right);\label{eq:a-final-sol}\\
u_{n} & =\mbox{i}\sqrt{k_{c}^{2}-\lambda_{n}^{2}}A_{n}\exp\left(\mbox{i}\sqrt{k_{c}^{2}-\lambda_{n}^{2}}x\right)\cos\left(\lambda_{n}\left(z+h\right)\right);\label{eq:u-final-sol}\\
p_{n} & =\mbox{i}\omega\rho A_{n}\exp\left(\mbox{i}\sqrt{k_{c}^{2}-\lambda_{n}^{2}}x\right)\cos\left(\lambda_{n}\left(z+h\right)\right);\label{eq:p-final-sol}
\end{alignat}
where $A_{n}$ is defined in equation \eqref{eq:A-Porous}.
\begin{figure}[H]
\begin{centering}
\includegraphics[width=0.5\textwidth]{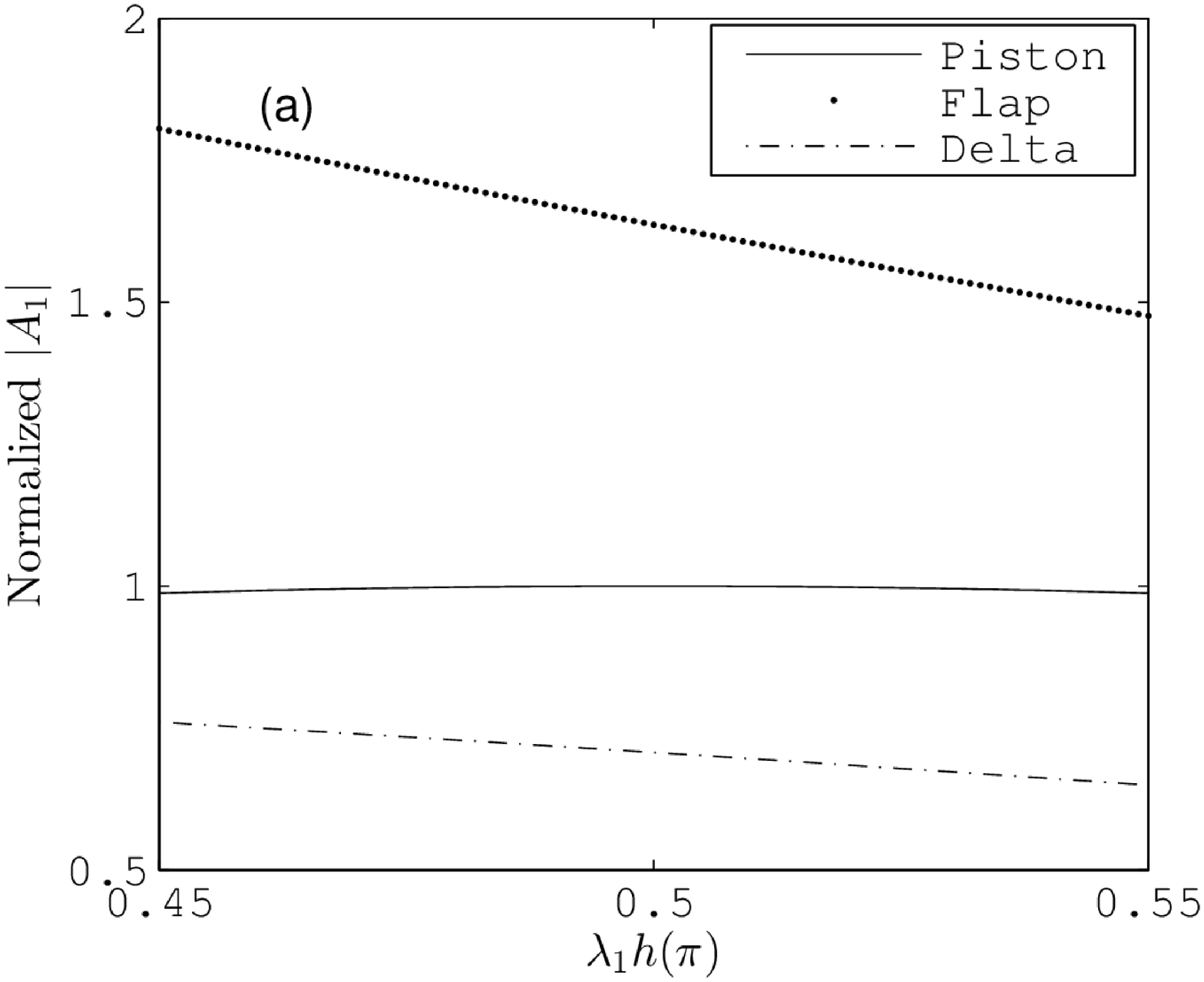}\includegraphics[width=0.5\textwidth]{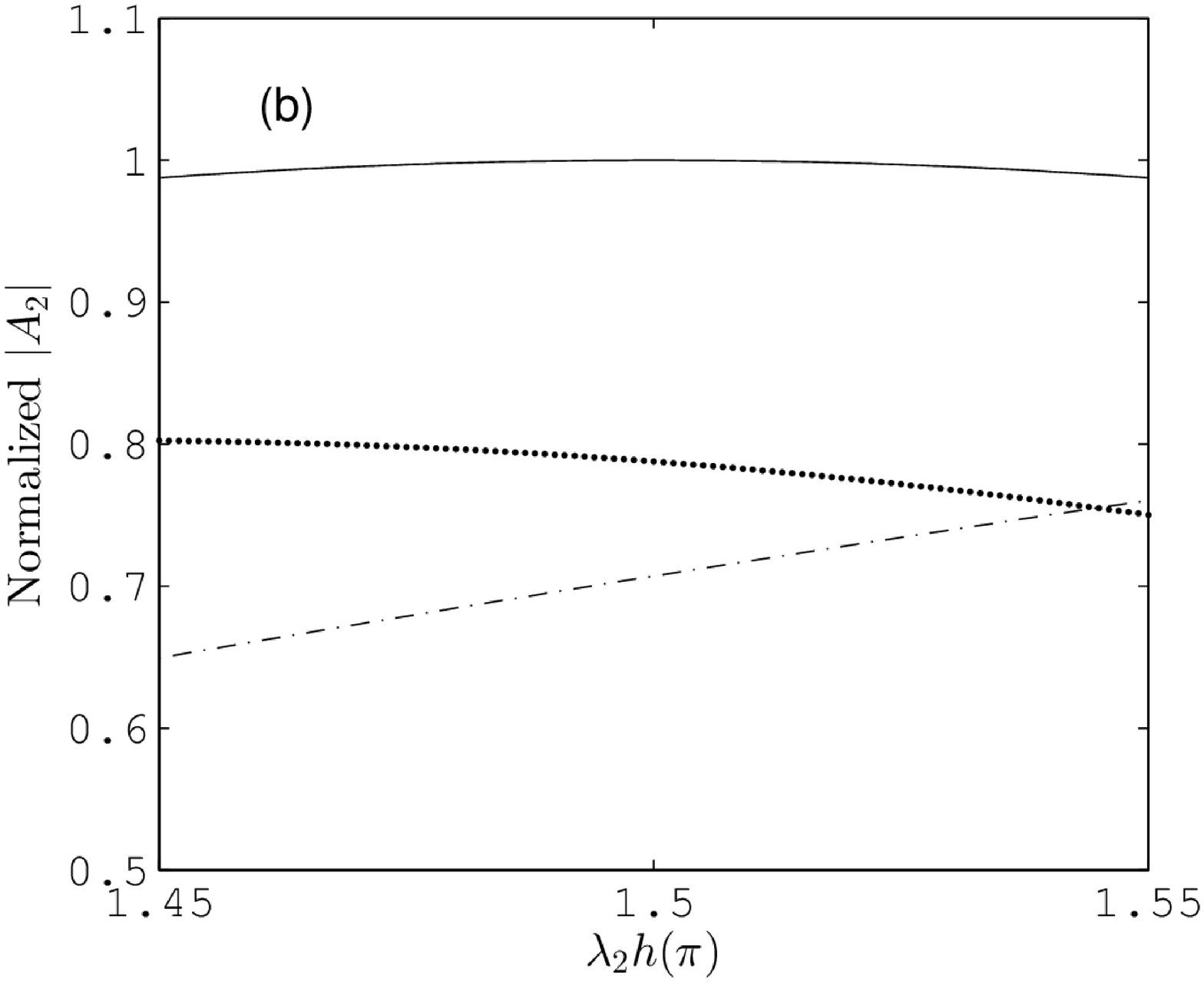}
\par\end{centering}
\centering{}\caption{\label{fig:Plates}Normalized AGW velocity potential amplitudes $\left|A_{1}\right|$
(a) and $\left|A_{2}\right|$ (b) by a factor of $-2\omega D/\lambda_{n}h\sqrt{k_{c}^{2}-\lambda_{n}^{2}}\left(1-CQ_{n}^{2}\right)$
($n=1$ and $2$, respectively) for three types of plate as a function
of $\lambda_{n}h$, which is chosen to vary between $[0.45,\:0.55]n\pi$.
The $\delta$-function type wavemaker is assumed to be located at
$z=-h/2$ for illustration purposes. Solid: piston-type plate; dotted:
flap-type plate; dashed: $\delta$-function type plate.}
\end{figure}

\subsection{Different types of plates}
We focus on the AGWs term in equations \eqref{eq:A-Havelock} and
\eqref{eq:A-Porous} and derive the explicit form based on different
types of the plate. Piston and flap motions \citep{Dean_Dalrymple_1991}
are commonly used for wave flumes in laboratory experiments, while
a wavemaker of $\delta$-function type is considered for deep ocean
\citep{Stuhlmeier_2015}. {Therefore the function
$d\left(z\right)$ that describes the piston motion in equation \eqref{eq:piston motion}
has been assumed to be}
\begin{equation}
d\left(z\right)=\left\{ \begin{array}{cc}
D, & \:\mbox{piston plate;}\\
D\left(1+z/h\right), & \:\mbox{flap plate;}\\
D\delta\left(z+h_{0}\right), & \:\mbox{\ensuremath{\delta}-function plate.}
\end{array}\right.\label{eq:plates}
\end{equation}

{Here, }$D$ is the horizontal amplitude of the stroke
motion. The $\delta$-function type wavemaker is located at $z=-h_{0}$.

Substituting equation \eqref{eq:plates} into the Havelock's solution
\eqref{eq:A-Havelock}, the expression for the amplitude of the velocity
potential, $A_{n}$, can be readily obtained 
\begin{alignat}{1}
A_{n} & =\left\{ \begin{array}{cc}
\frac{-2\omega D}{\lambda_{n}h\sqrt{k_{c}^{2}-\lambda_{n}^{2}}\left(1-CQ_{n}^{2}\right)}\sin\left(\lambda_{n}h\right), & \:\mbox{piston plate;}\\
\frac{-2\omega D}{\lambda_{n}h\sqrt{k_{c}^{2}-\lambda_{n}^{2}}\left(1-CQ_{n}^{2}\right)}\frac{1}{\lambda_{n}h}\left[\lambda_{n}h\sin\left(\lambda_{n}h\right)+\cos\left(\lambda_{n}h\right)-1\right], & \:\mbox{flap plate;}\\
\frac{-2\omega D}{\lambda_{n}h\sqrt{k_{c}^{2}-\lambda_{n}^{2}}\left(1-CQ_{n}^{2}\right)}\cos\left(\lambda_{n}\left(h-h_{0}\right)\right), & \:\mbox{\ensuremath{\delta}-function plate.}
\end{array}\right.\label{eq:A}
\end{alignat}

A comparison of the normalized velocity potential amplitude in equation
\eqref{eq:A} is given in Figure \ref{fig:Plates}. Apparently, a
flap wavemaker produces the largest first-mode AGW for the same plate-motion
amplitude $D$. For a flap plate, equation \eqref{eq:A} also indicates
that the normalized AGW amplitudes are inversely proportional to $\lambda_{n}h$,
therefore higher AGW modes must have smaller normalized amplitudes.

\section{Examples \label{sec:Examples}}

\subsection{Acoustic-gravity waves in a wave flume}

The number of existing AGW modes associated with specific frequency
and depth can be {calculated} by equation \eqref{eq:dispersion-gravity}.
This relationship shows that more AGW modes can be generated at a
higher frequency $\omega$ or deeper water. Therefore{, creating AGWs in the laboratory is not an easy task.  In order to obey AGW theory (with the absence of bottom elasticity, e.g. see \cite{Eyov_2013} for the detailed analysis) and create AGWs in the laboratory experiments we need to operate at relatively very high frequencies.} For example, there are three AGW modes corresponding
to a $5$ kHz wavemaker in a $0.5$ m wave flume. {Although, working with a $5$ kHz introduces some real difficulties, we can still have feasible experiments with piezoelectric membranes to validate the proposed theory (an undergoing research). Alternatively, one needs to carry out an experiment in the deep ocean, which is by no means easier to perform. Due to this conflicting choice of experimental environment, we dedicate this section and the following to the disparate wave flume and deep ocean systems, respectively. } 

\begin{figure}[H]
\begin{centering}
\includegraphics[width=0.5\textwidth]{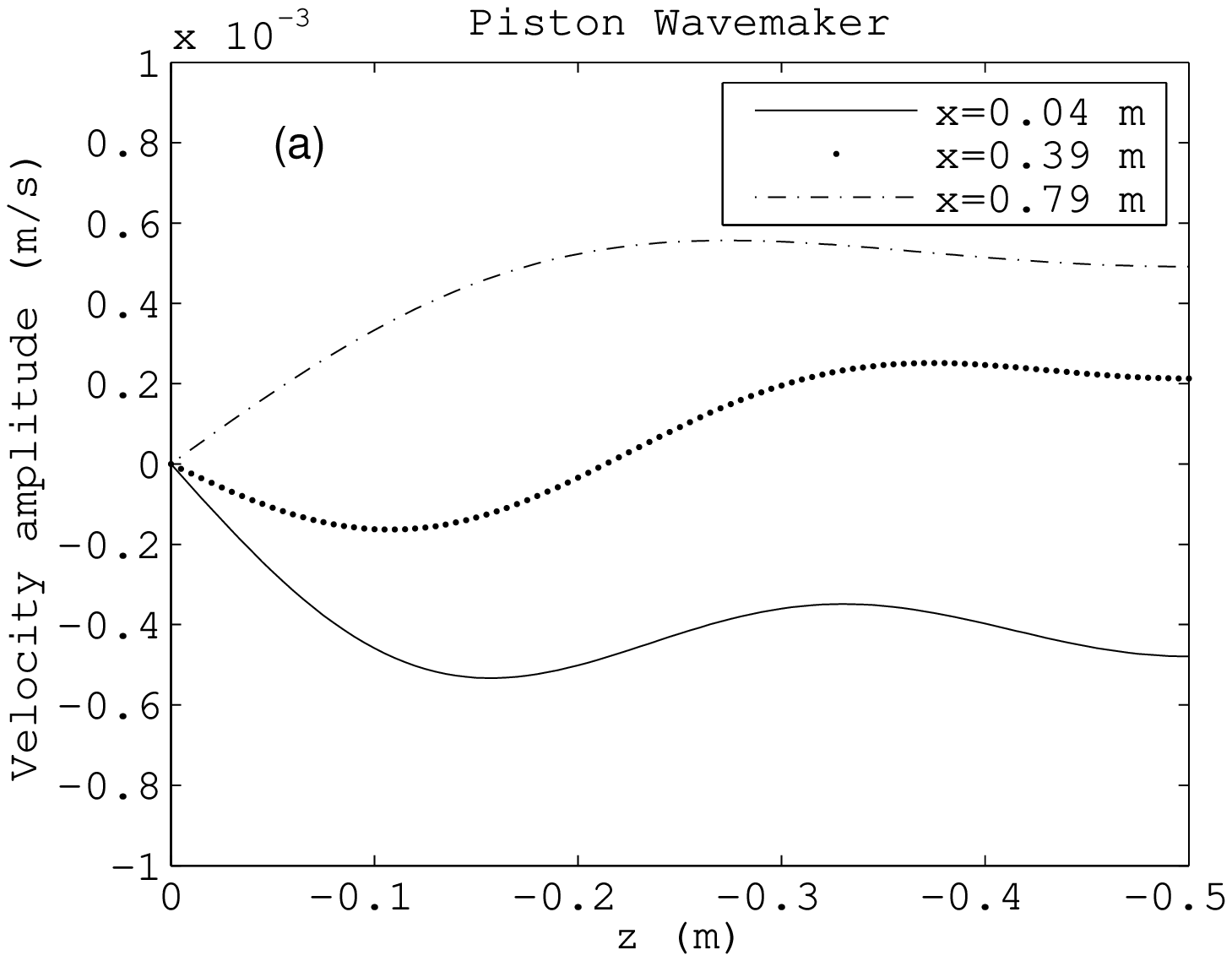}\includegraphics[width=0.5\textwidth]{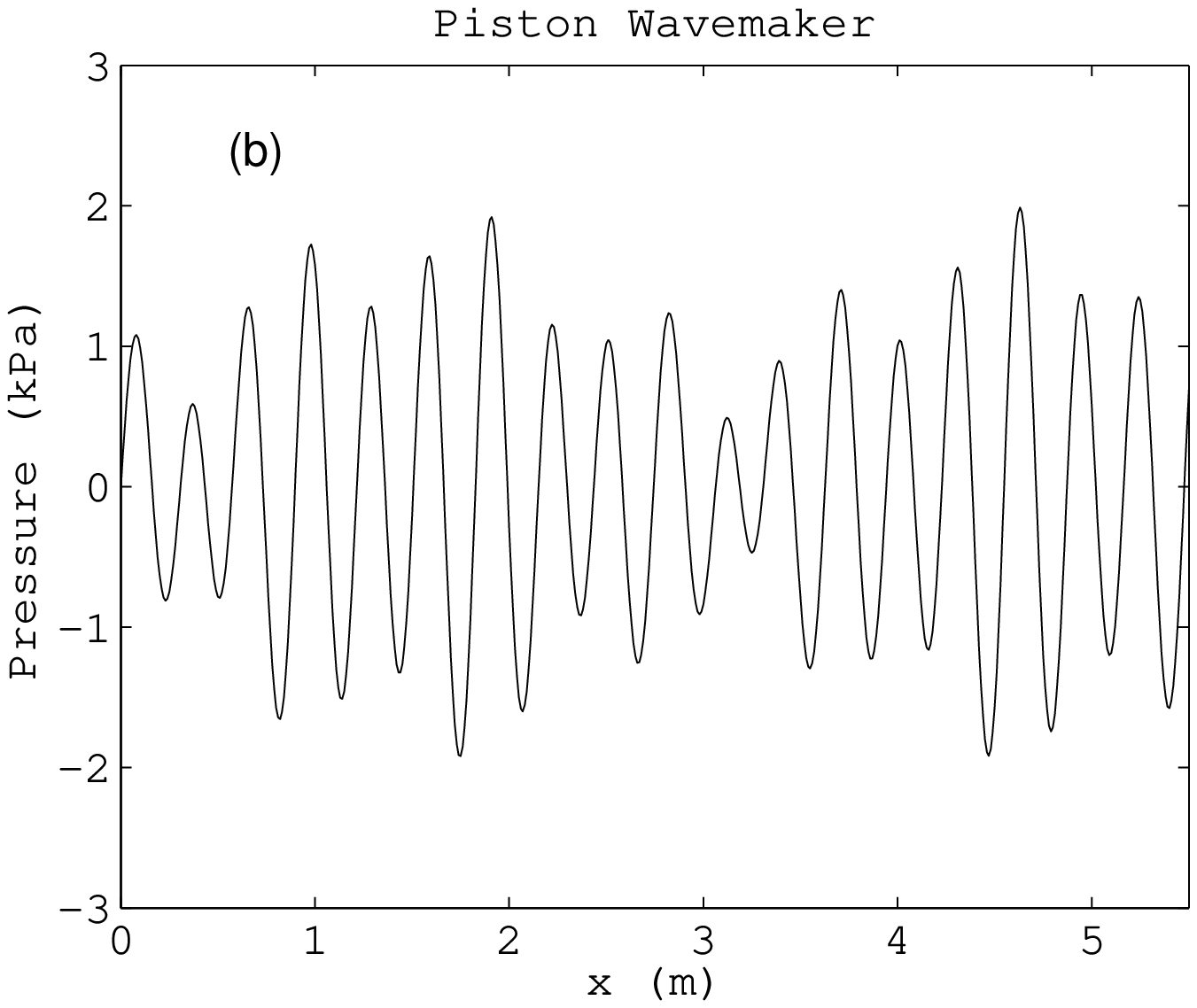}
\par\end{centering}
\begin{centering}
\includegraphics[width=0.5\textwidth]{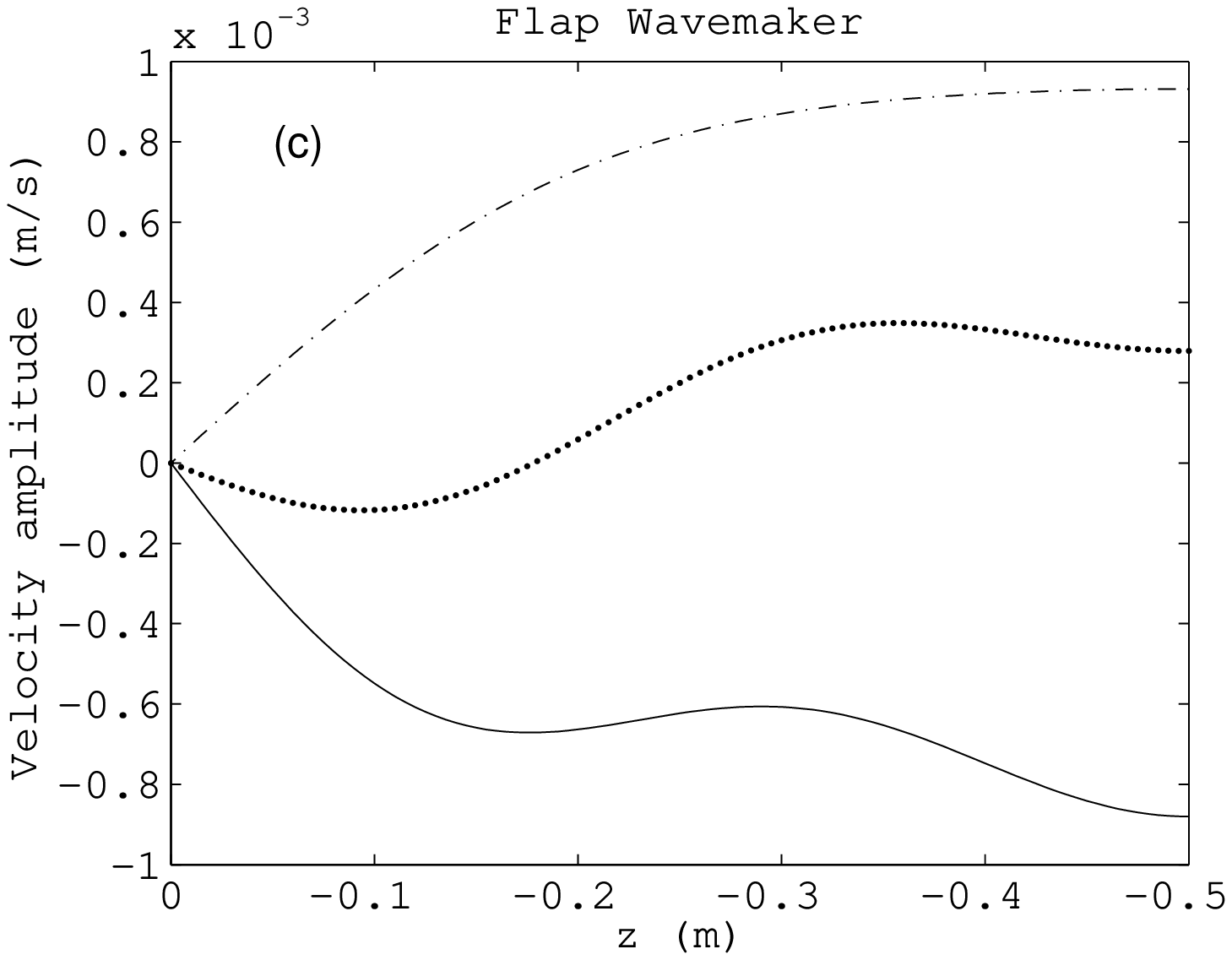}\includegraphics[width=0.5\textwidth]{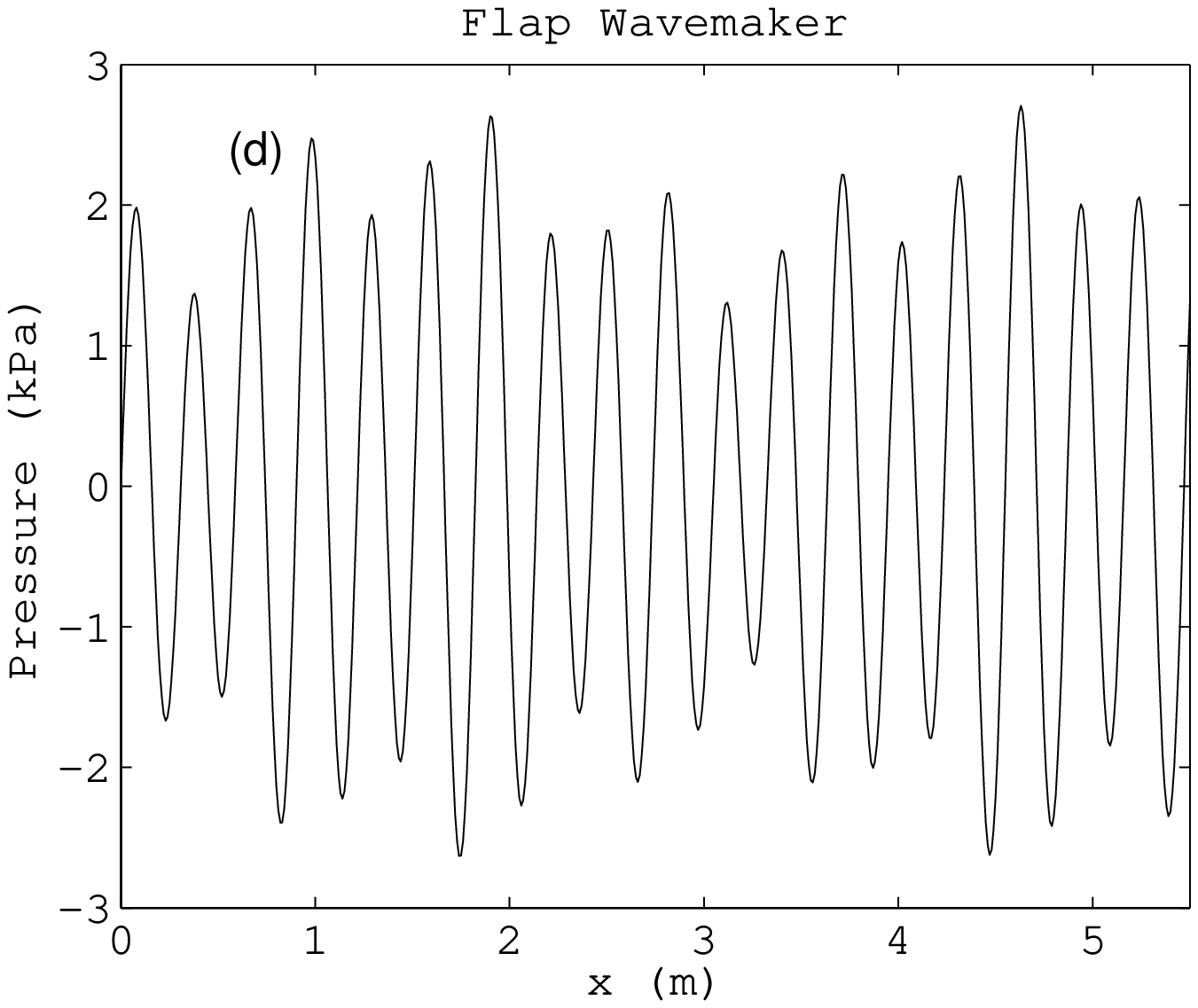}
\par\end{centering}
\centering{}\caption{\label{fig:Havelock-Piston}Acoustic-gravity waves generated by a
wavemaker in a wave flume for $f=5$ kHz based on the Havelock's wavemaker
theory; $L=5.5$ m, $h=0.5$ m, $b=1$ m, $c=1500$ m/s. The motion
of the plate is limited by the constraint that, the horizontal movement
$\leq2.1$ m, the horizontal velocity $\leq3.8$ m/s, horizontal acceleration
$\leq19.6$ m/s$^{2}$ (parameters come from the unidirectional wavemaker
in O.H. Hinsdale Wave Research Laboratory, Oregon State University).
(a), (c) Vertical distribution of the velocity amplitude at $x=0.04$
m, $0.39$ m, and $0.79$ m away from the wavemaker. (b), (d) Horizontal
distribution of the pressure amplitude at the bottom of the flume.
(a), (b) piston wavemaker; (c), (d) flap wavemaker.}
\end{figure}

\begin{figure}[H]
\begin{centering}
\includegraphics[width=0.5\textwidth]{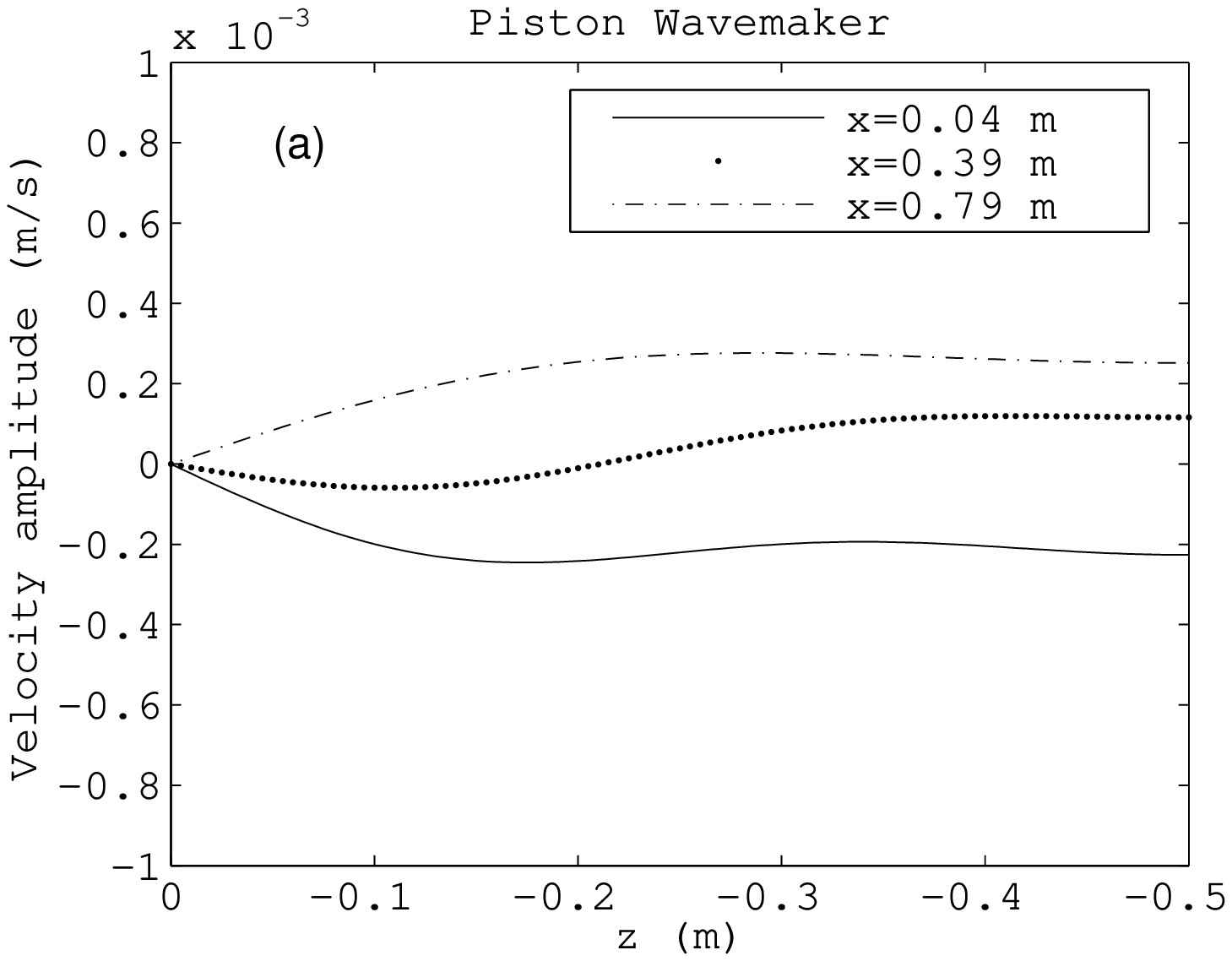}\includegraphics[width=0.5\textwidth]{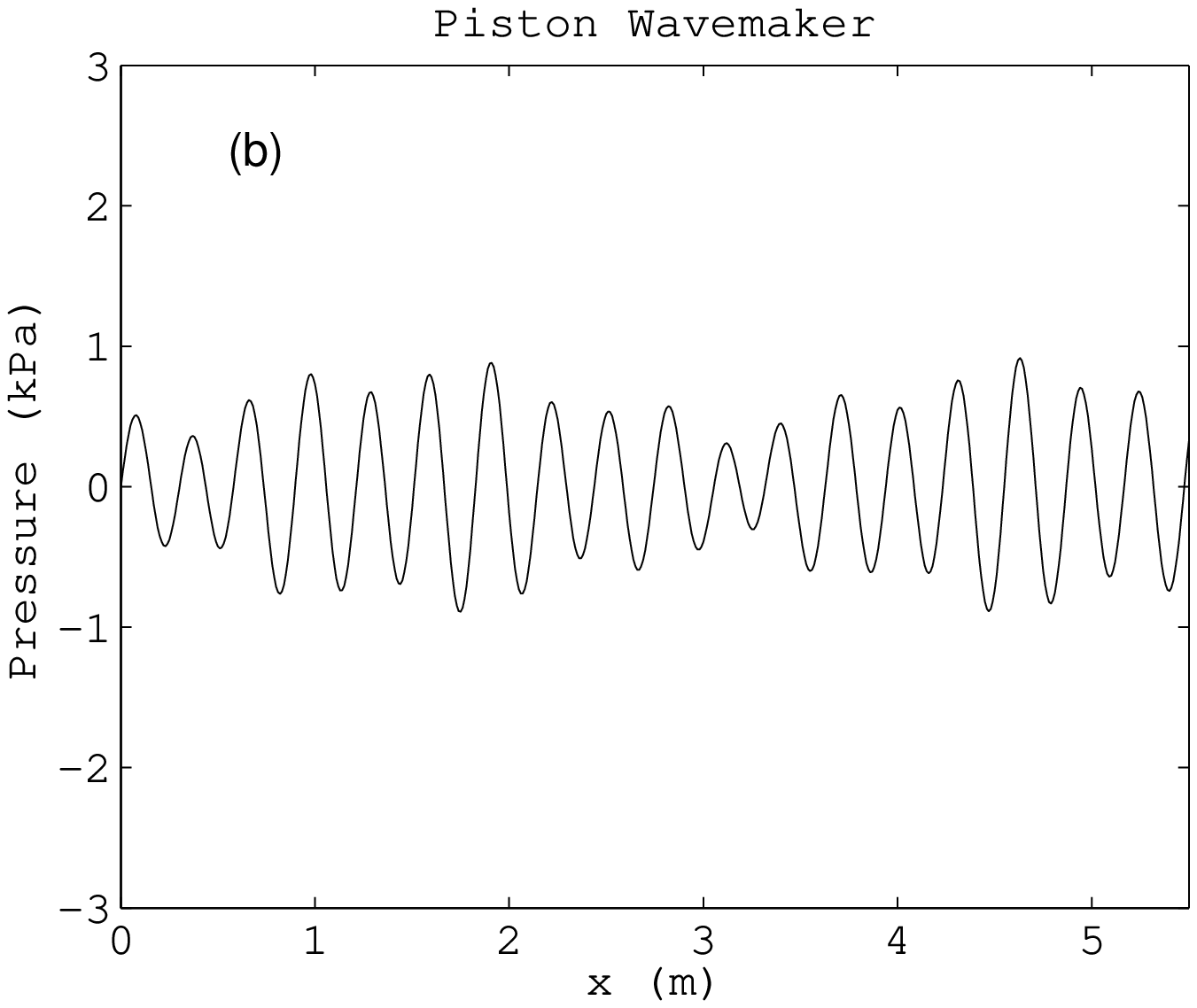}
\par\end{centering}
\begin{centering}
\includegraphics[width=0.5\textwidth]{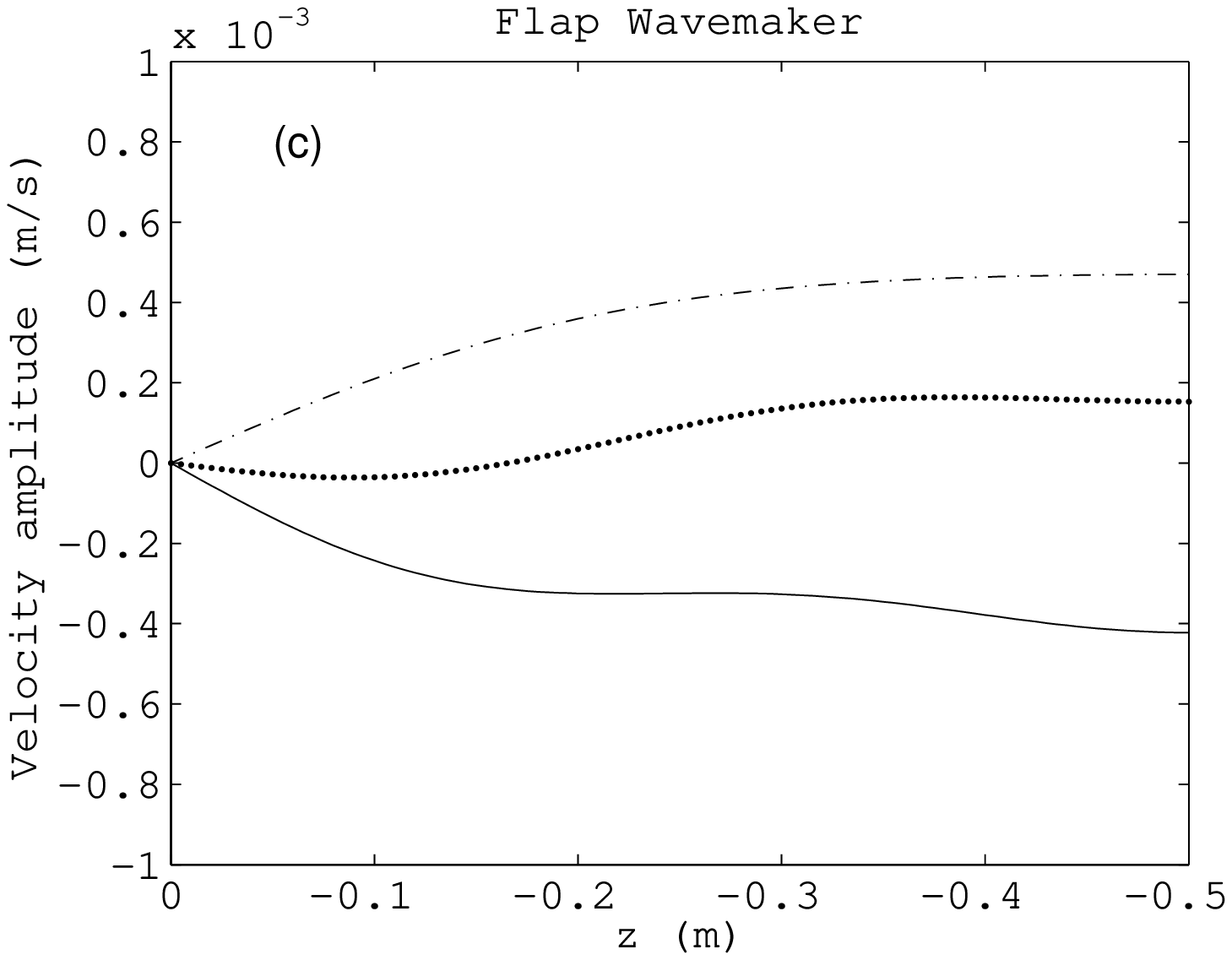}\includegraphics[width=0.5\textwidth]{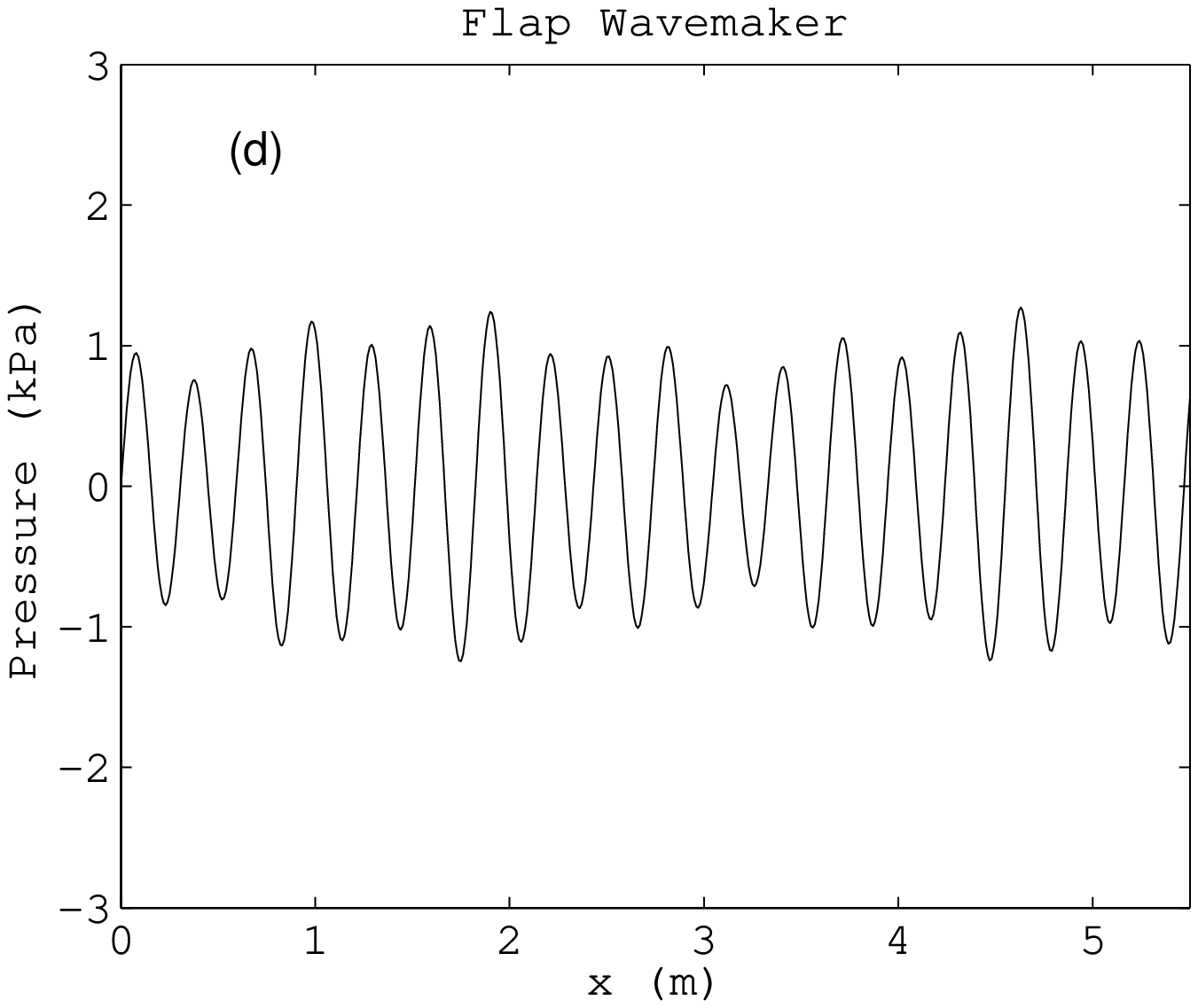}
\par\end{centering}
\centering{}\caption{\label{fig:Porous-Piston}Acoustic-gravity waves generated by a wavemaker
in a wave flume for $f=5$ kHz based on the porous wavemaker theory;
$L=5.5$ m, $h=0.5$ m, $b=1$ m, $c=1500$ m/s. The motion of the
plate is limited by the constraint that, the horizontal movement $\leq2.1$
m, the horizontal velocity $\leq3.8$ m/s, horizontal acceleration
$\leq19.6$ m/s$^{2}$ (parameters come from the unidirectional wavemaker
in O.H. Hinsdale Wave Research Laboratory, Oregon State University).
(a), (c) Vertical distribution of the velocity amplitude at $x=0.04$
m, $0.39$ m, and $0.79$ m away from the wavemaker. (b), (d) Horizontal
distribution of the pressure amplitude at the bottom of the flume.
(a), (b) piston wavemaker; (c), (d) flap wavemaker.}
\end{figure}

Examples of AGWs
generated in a wave flume by piston- and flap-type plates are shown
in Figures \ref{fig:Havelock-Piston} and \ref{fig:Porous-Piston}
based on Havelock's and porous wavemaker theories, respectively. Notice
that the stroke motion is not only limited by its maximum stroke distance,
but also the maximum velocity and acceleration. Here, we assume the
wavemaker has the same constraints as the unidirectional wavemaker
of the O.H. Hinsdale Wave Research Laboratory in Oregon State University
(for example the laboratory experiment presented in \citealp{Tian_etal_2015});
simple calculation using equations \eqref{eq:piston-velocity-acceleration}
shows that the stroke amplitude of a wavemaker with $f=5$ kHz is
at the order of $10^{-8}$ m, {which requires a very careful experiment.}

It is difficult to measure the AGWs surface elevations directly because
of their small amplitudes (not shown in the figure), {although}
the horizontal velocity {component} is at the order
of $10^{-3}$ m/s (Figure \ref{fig:Havelock-Piston} (a) and (c)),
which is detectable using a particle image velocimetry (PIV) system.
It is also worth mentioning that, unlike gravity waves,
AGWs velocity amplitude oscillates vertically and leaves a distinct
pressure signature throughout the entire water column{, and in particular at the bottom. }

The time series of the pressure at the bottom (Figure \ref{fig:Havelock-Piston}
(b) and (d)) behaves similarly as that of the surface elevation, although
measurable by a wired pressure sensor. Therefore, in spite of the
small amplitude of their surface elevation, AGWs are expected to be
detectable on bottom-pressure records or PIV velocity measurement
in a laboratory study. On the other hand the flap wavemaker is able
to produce larger waves compared to a piston one as shown in Figure
\ref{fig:Havelock-Piston} (c) and (d). Figure \ref{fig:Porous-Piston}
presents AGWs-induced pressure and velocities produced by porous wavemakers.
Due to porosity effects, the velocity and pressure amplitudes are
generally smaller than these in Havelock's theory.

\subsection{Acoustic-gravity waves in deep ocean}

\begin{figure}[H]
\begin{centering}
\includegraphics[width=0.5\textwidth]{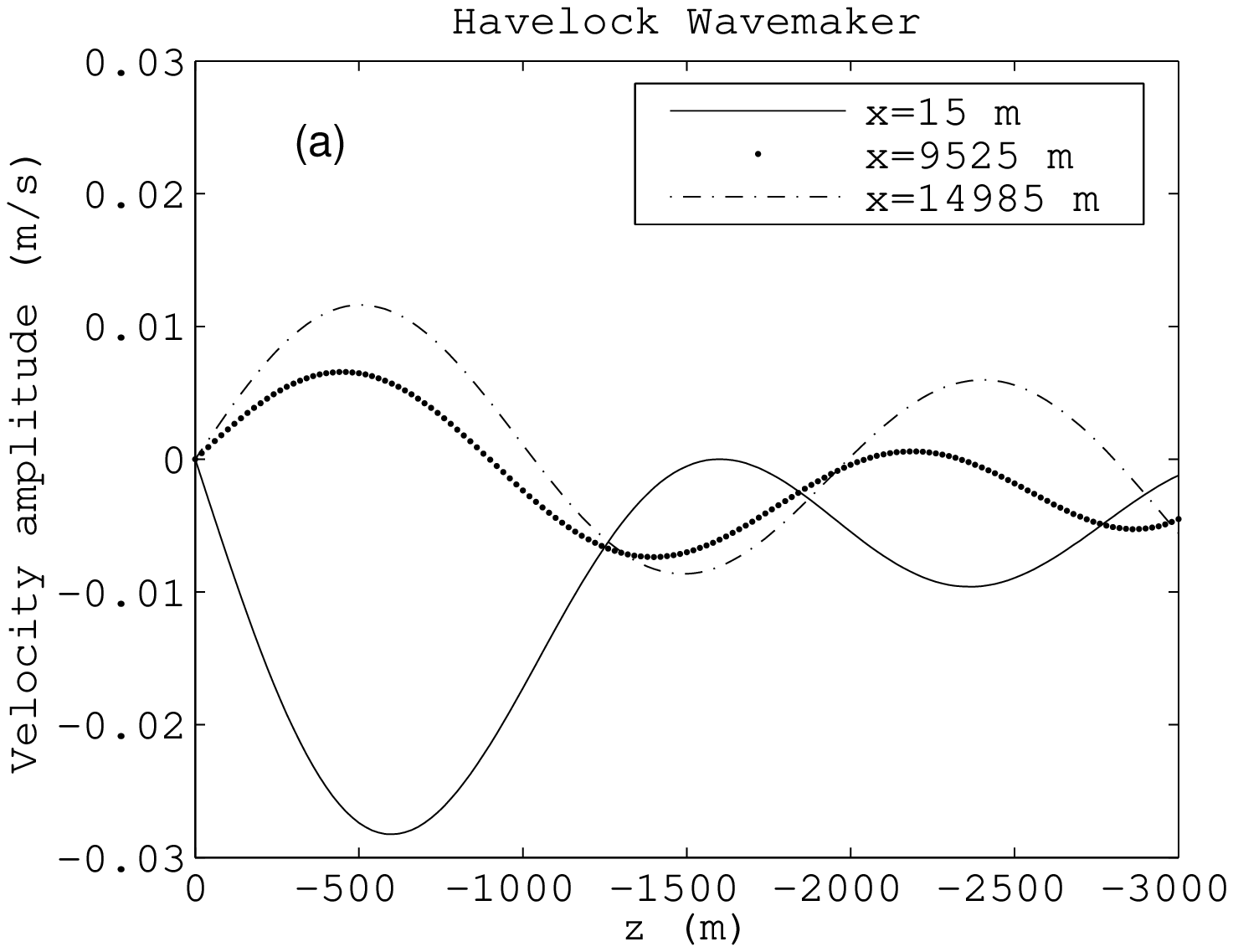}\includegraphics[width=0.5\textwidth]{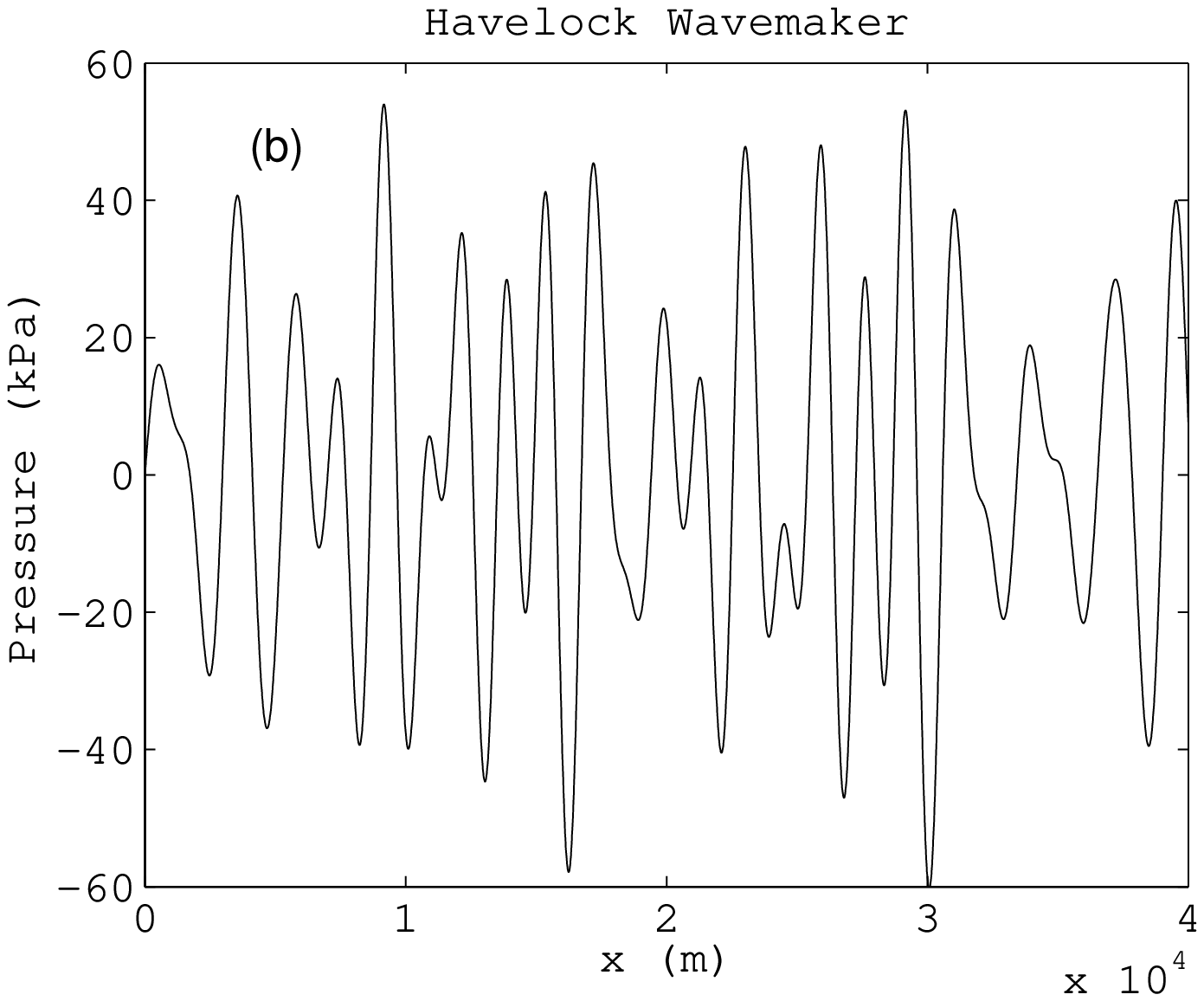}
\par\end{centering}
\begin{centering}
\includegraphics[width=0.5\textwidth]{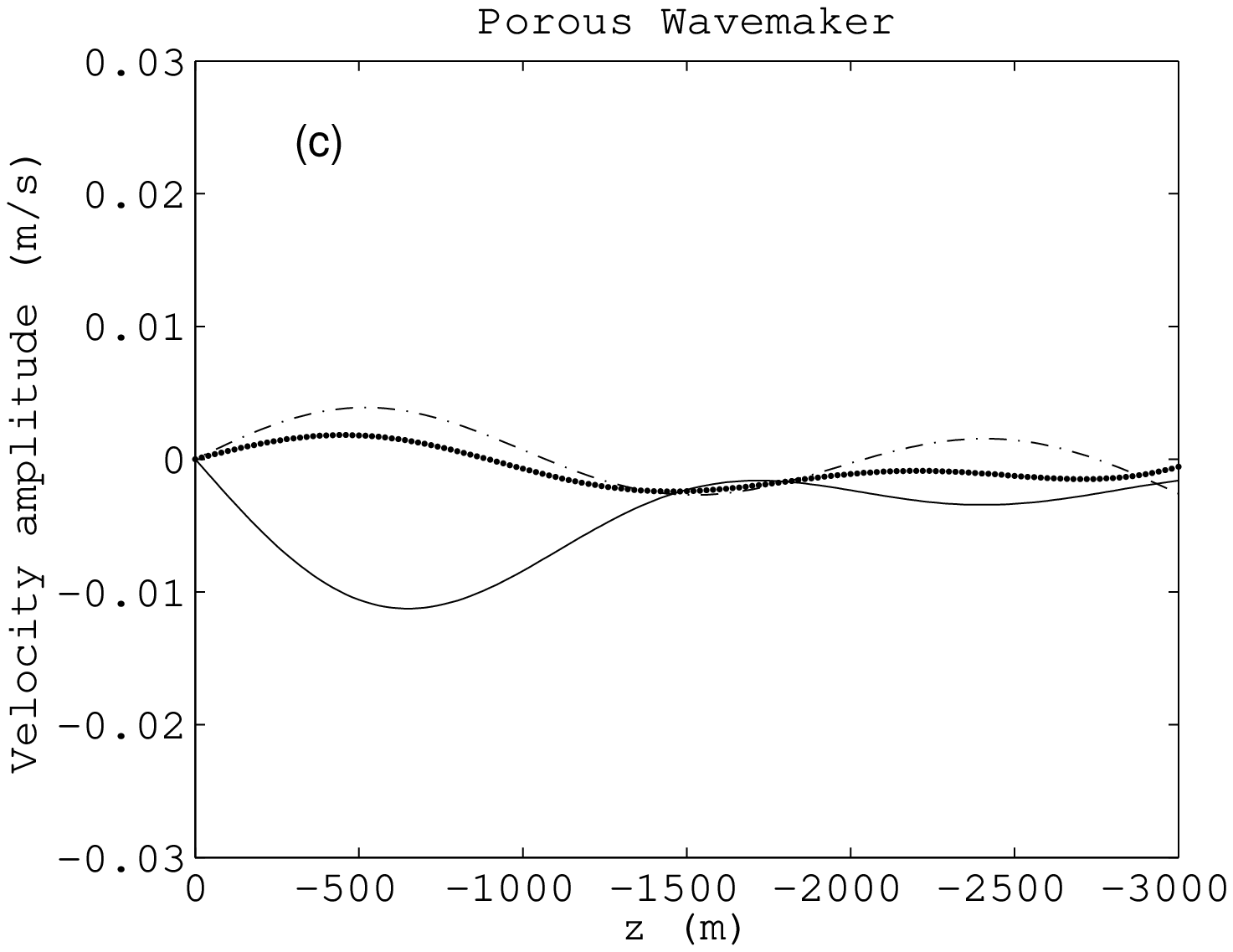}\includegraphics[width=0.5\textwidth]{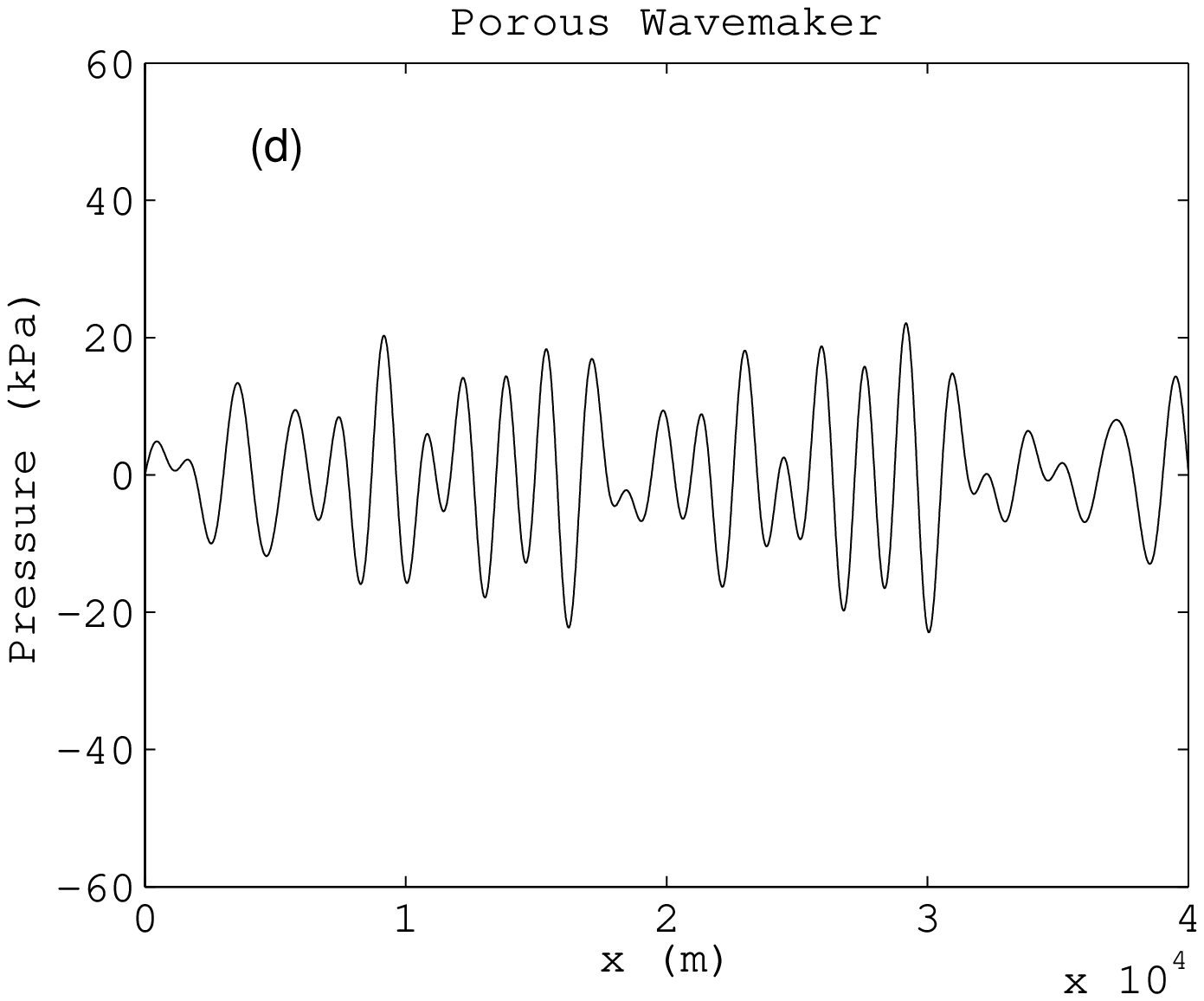}
\par\end{centering}
\centering{}\caption{\label{fig:Havelock-Delta}Acoustic-gravity waves generated by a $\delta$-function
wavemaker in the ocean placed at $z=-12.5$ m for $f=1$ Hz. The ocean
is $h=4000$ m deep and the speed of sound is $c=1500$ m/s. (a),
(c) Vertical distribution of the velocity amplitude at $x=15$ m,
$9525$ m, and $14985$ m away from the wavemaker. (b), (d) Horizontal
distribution of the pressure amplitude at the bottom of the flume.
(a), (b) Havelock's wavemaker theory; (c), (d) porous wavemaker theory.}
\end{figure}

{We treat the problem of a wavemaker plate in deep
ocean as a point source in deep water (similar to the ocean acoustics
problem in \citealt{Jensen-book}) , and consider it as a $\delta$-function.}
An example of a $\delta$-function wavemaker located at $z=-12.5$
m in deep ocean with $f=1$ Hz and $h=4000$ m is presented in Figure
\ref{fig:Havelock-Delta} for both Havelock's and porous wavemaker
theories. The AGWs produced by an impermeable wavemaker ($G=0$) have
larger amplitudes (Figure \ref{fig:Havelock-Delta} (a) and (b)),
whereas the wave amplitude decreases as the wavemaker becomes porous
($G\neq0$). The surface elevation increases to $10^{-3}$ m (not
shown) compared to those in the laboratory example, although still
hard to be distinguished from that of surface gravity waves. The velocity
amplitudes are almost zero at the surface; {they reach their maximum at
about $z=500$ m, and oscillate across the water column in the $z$-direction}. Although AGWs
have similar frequencies as the gravity mode, their distributions
are periodic throughout the water column {(i.e. do not decay with depth)}, therefore can be distinguished
from the decaying behavior of gravity waves. The order of magnitude
of the velocity reaches $10^{-2}$ m/s, which is measurable by standard instruments such as the ADCP (Acoustic Doppler Current Profiler),
PCADP (Pulse-coherent Acoustic Doppler Profiler). It is thus suggested
to make use of pressure sensors at the seabed, or deep below the surface, where surface-wave
signatures are negligible. The AGWs signal, however, is at the order
of $10$ kPa, which is easy to measure (e.g., the MODE experiment
that measures the pressure fluctuation on the deep-sea floor by \citealp{Brown_JPO1075}).
This simple example shows that AGWs may be responsible for the low-frequency
oceanic noise on the seabed \citep[e.g.,][]{Ardhuin_etal_2013}.

\section{Conclusion \label{sec:Conclusion}}

Without the usual overlook of the slight compressibility of water,
we present Havelock's and porous wavemaker theories to analyze different
modes of water waves following \citet{Dean_Dalrymple_1991} and \citet{Chwang_JFM1983},
with special focus on understanding the AGW mode. These
theories may have important {implications} in the study of surface waves
in flume experiment \citep{Stuhlmeier_2015} or tsunamis caused by
landslides during earthquakes in deep ocean \citep{Yamamoto_1982,Stiassnie_2010,Kadri_Stiassnie_2012}.
Moreover, the generation of acoustic-gravity waves can be attributed
to wave-structure interaction \citep{Stuhlmeier_2015}, therefore
another possible implication is where the efficiency of
wave-energy harnessing devices is of interest, with the wavemaker
being subjected to some form of wave energy converter, e.g., a flap
gate \citep[e.g.,][]{Sammarco_2013}. {Another and probably a more immediate implication is the remote detection of the wavy sea-state which can help tuning the surface wave energy converters for maximum efficiency. These are left for future study, and we hope this work will motivate scientists and engineers to look into these important implications.}

Both Havelock's and porous wavemaker solutions reduce to previous
theories \citep{Dean_Dalrymple_1991,Chwang_JFM1983} for incompressible
flow when the compressibility coefficient $k_{c}$ in equation \eqref{eq:Helmholtz}
tends to zero. The solutions for three types of plates as well as
the spatial distribution of the AGWs components are presented. Having
the same horizontal displacements of the plate, the results suggest
that a flap wavemaker is capable of making lager waves than piston
and $\delta$-function-type wavemakers. The spatial distribution of
the amplitude of the surface elevation, horizontal velocity, and bottom
pressure due to both theories was compared. It is shown that, generally, the porous
wavemaker results in smaller waves than those produced by
Havelock's theory due to the porosity factor $G_{n}$ in equation
\eqref{eq:porous factor}. The calculations reveal that the surface
elevation of AGWs in the current lab experimental settings is at the
order of $10^{-9}$ m, and can reach $10^{-3}$ m in deep ocean. Therefore
surface elevation of AGWs is hard to measure; while velocity amplitude
suggests that AGWs can be detected by a particle image velocimetry
(PIV) system in the laboratory experiment. Finally, the pressure distributions
show that AGW signals are significantly large at the bottom of a
wave flume and deep ocean, to be captured by a standard pressure sensor.
This study motivates further laboratory studies and field measurement
on deep ocean as it predicts the characteristics of these waves, thus
provides insights on how to carry out direct measurements. It also
sheds some light on development of tsunami early-detection systems
from the perspective of describing AGWs near the epicenter when earthquakes
occur. Finally, the porous wavemaker theory can potentially contribute
to the the study of deep-ocean energy-harvest device where the porous
plates can be treated as an energy absorber.

\subsection*{Acknowledgment}

The first author acknowledges the Postdoctoral Fellowship at
the University of Haifa in collaboration with Woods Hole Oceanographic
Institution, part of which this work took place.

\end{document}